\documentclass[
 singlecolumn,
groupedaddress,
 amsmath,amssymb,
 aps,
pra,
]{revtex4-1}

\usepackage[utf8]{inputenc}
\usepackage{siunitx}
\usepackage{amsmath}
\usepackage{cleveref}
\usepackage{tikz}
\usetikzlibrary{calc}
\usepackage{pgfplots}
\usepackage{braket}
\usepackage{nicefrac}

\usepackage{todonotes}

\usepackage{siunitx}
\usepackage{amsmath}
\usepackage{cleveref}
\usepackage{braket}
\usepackage{nicefrac}
\usepackage{todonotes}
\usepackage{pgfplots}
\usepackage{pgf}
\usepackage{tikz}

\usetikzlibrary{scopes}         % Shorthand for Scope Environments
\usetikzlibrary{decorations.pathmorphing}
\usetikzlibrary{intersections}  % Intersections of Arbitrary Paths
\usetikzlibrary{calc}           % Coordinate Calculations
\usetikzlibrary{positioning}    % Advanced Placement Options
\usetikzlibrary{arrows}
\usetikzlibrary{backgrounds}
\usetikzlibrary{fit}
\usetikzlibrary{fixedpointarithmetic}
\usetikzlibrary{fpu}
\usetikzlibrary{patterns}
\usetikzlibrary{plothandlers}
\usetikzlibrary{plotmarks}
\usetikzlibrary{shadings}
\usetikzlibrary{topaths}
\usepackage{xspace}

\newcommand{\Eotvos}{\name{E\"otv\"os}}

\newcommand{\keff}{\ensuremath{k_{\text{eff}}}\xspace}
\newcommand{\keffi}[1]{k_{\text{eff,#1}}}

\newcommand{\name}[1]{#1}

\newcommand{\quot}[1]{``#1''}

\newcommand{\stat}{5.4\times10^{-7}}
\newcommand{\etav}{(0.3\pm 5.4)\times 10^{-7}}

\newcommand{\Rb}{\textsuperscript{87}Rb\xspace}
\newcommand{\K}{\textsuperscript{39}K\xspace}

\newcommand{\Yb}[1]{\textsuperscript{#1}Yb\xspace}

\newcommand{\Eot}{$\eta_{\,\text{Rb,K}}$}

\newcommand{\EotDelta}{$\Delta\eta$}
\newcommand{\Eotsigma}{$\delta\eta$}

\definecolor{MyRed}{HTML}{FF5555}
\definecolor{MyDarkRed}{HTML}{B6321C}
\definecolor{MyBlue}{HTML}{0071BC} %wie RoyalBlue
\definecolor{MyGreen}{HTML}{009B55} %wie ForestGreen
\definecolor{MyOrange}{HTML}{FFB368}

\pgfplotscreateplotcyclelist{mycycle}{%
black,only marks,mark size=\mymarker,every mark/.append style={fill=black},mark=square*\\%
black,\myline\\%
MyRed,only marks,mark size=1.2*\mymarker,every mark/.append style={fill=MyRed},mark=*\\%
MyRed,dashed,\myline\\%
MyBlue,only marks,mark size=1.5*\mymarker,every mark/.append style={fill=MyBlue},mark=diamond*\\%
MyBlue,dotted,\myline\\%
MyGreen,only marks,mark size=\mymarker,every mark/.append style={fill=MyGreen},mark=otimes*\\%
MyGreen,dashdotted,\myline\\%
}

\pgfplotscreateplotcyclelist{mycycle2}{%
black,only marks,mark size=0.4*\mymarker,every mark/.append style={fill=black},mark=square*\\%
black\\%
MyRed,only marks,mark size=1.2*0.4*\mymarker,every mark/.append style={fill=MyRed},mark=*\\%
MyRed,dashed\\%
MyBlue,only marks,mark size=1.5*0.4*\mymarker,every mark/.append style={fill=MyBlue},mark=diamond*\\%
MyBlue,dotted\\%
MyGreen,only marks,mark size=0.4*\mymarker,every mark/.append style={fill=MyGreen},mark=otimes*\\%
MyGreen,dashdotted\\%
}

\pgfplotscreateplotcyclelist{mycycle2b}{%
MyRed,only marks,mark size=1.2*0.4*\mymarker,every mark/.append style={fill=MyRed},mark=*\\%
MyRed,dashed\\%
MyBlue,only marks,mark size=1.5*0.4*\mymarker,every mark/.append style={fill=MyBlue},mark=diamond*\\%
MyBlue,dotted\\%
MyGreen,only marks,mark size=0.4*\mymarker,every mark/.append style={fill=MyGreen},mark=otimes*\\%
MyGreen,dashdotted\\%
}

\pgfplotscreateplotcyclelist{mycycle2c}{%
MyBlue,only marks,mark size=1.5*0.4*\mymarker,every mark/.append style={fill=MyBlue},mark=diamond*\\%
MyBlue,dotted\\%
MyGreen,only marks,mark size=0.4*\mymarker,every mark/.append style={fill=MyGreen},mark=otimes*\\%
MyGreen,dashdotted\\%
}

\pgfplotscreateplotcyclelist{mycycle3}{%
black,only marks,mark size=0.4*\mymarker,every mark/.append style={fill=black},mark=square*\\%
black\\%
MyRed,dashed,\myline\\%
MyBlue,only marks,mark size=1.5*0.4*\mymarker,every mark/.append style={fill=MyBlue},mark=diamond*\\%
MyBlue,dotted\\%
MyGreen,only marks,mark size=0.4*\mymarker,every mark/.append style={fill=MyGreen},mark=otimes*\\%
MyGreen,dashdotted\\%
}

\pgfplotscreateplotcyclelist{mycycle4}{%
black,only marks,mark size=\mymarker,every mark/.append style={fill=black},mark=square*\\%
black\\%
MyRed,dashed,\myline\\%
MyBlue,only marks,mark size=1.5*\mymarker,every mark/.append style={fill=MyBlue},mark=diamond*\\%
MyBlue,dotted\\%
MyGreen,only marks,mark size=\mymarker,every mark/.append style={fill=MyGreen},mark=otimes*\\%
MyGreen,dashdotted\\%
}

\pgfplotscreateplotcyclelist{mycycleins}{%
MyRed,only marks,mark size=1.2*\mymarker,every mark/.append style={fill=MyRed},mark=*\\%
MyRed,dashed,\myline\\%
}

\pgfplotscreateplotcyclelist{mycyclesim}{%
black,only marks,mark size=\mymarker,every mark/.append style={fill=black},mark=square*\\%
black,only marks,mark size=0.4*\mymarker,every mark/.append style={fill=black},mark=square*\\%
black,only marks,mark size=0.4*\mymarker,every mark/.append style={fill=black},mark=square*\\%
black,\myline\\%
MyRed,only marks,mark size=1.2*\mymarker,every mark/.append style={fill=MyRed},mark=*\\%
MyRed,only marks,mark size=1.2*0.4*\mymarker,every mark/.append style={fill=MyRed},mark=*\\%
MyRed,only marks,mark size=1.2*0.4*\mymarker,every mark/.append style={fill=MyRed},mark=*\\%
MyRed,dashed,\myline\\%
MyBlue,only marks,mark size=1.5*\mymarker,every mark/.append style={fill=MyBlue},mark=diamond*\\%
MyBlue,dotted,\myline\\%
MyGreen,only marks,mark size=\mymarker,every mark/.append style={fill=MyGreen},mark=otimes*\\%
MyGreen,dashdotted,\myline\\%
}

\pgfplotscreateplotcyclelist{mylinecycle}{%
black,\myline\\%
MyRed,dashed,\myline\\%
MyBlue,dotted,\myline\\%
MyGreen,dashdotted,\myline\\%
}

\pgfplotscreateplotcyclelist{mylinecycle2}{%
MyRed,dashed,\myline\\%
MyBlue,dotted,\myline\\%
MyGreen,dashdotted,\myline\\%
}

\pgfplotscreateplotcyclelist{mydetectioncycle}{%
MyRed,dashed,\myline\\%
MyBlue,dotted,\myline\\%
MyGreen,dashdotted,\myline\\%
MyOrange,loosely dashed,\myline\\%
}

\pgfplotscreateplotcyclelist{myadevcycle}{%
black,only marks,mark size=\mymarker,every mark/.append style={fill=black},mark=square*\\%
black,only marks,mark size=0.4*\mymarker,every mark/.append style={fill=black},mark=square*\\%
black,only marks,mark size=0.4*\mymarker,every mark/.append style={fill=black},mark=square*\\%
MyRed,only marks,mark size=1.2*\mymarker,every mark/.append style={fill=MyRed},mark=*\\%
MyRed,only marks,mark size=1.2*0.4*\mymarker,every mark/.append style={fill=MyRed},mark=*\\%
MyRed,only marks,mark size=1.2*0.4*\mymarker,every mark/.append style={fill=MyRed},mark=*\\%
MyBlue,only marks,mark size=1.5*\mymarker,every mark/.append style={fill=MyBlue},mark=diamond*\\%
MyBlue,only marks,mark size=1.5*0.4*\mymarker,every mark/.append style={fill=MyBlue},mark=diamond*\\%
MyBlue,only marks,mark size=1.5*0.4*\mymarker,every mark/.append style={fill=MyBlue},mark=diamond*\\%
}

\pgfplotscreateplotcyclelist{myadevcycle2}{%
MyBlue,only marks,mark size=1.5*\mymarker,every mark/.append style={fill=MyBlue},mark=diamond*\\%
MyBlue,only marks,mark size=1.5*0.4*\mymarker,every mark/.append style={fill=MyBlue},mark=diamond*\\%
MyBlue,only marks,mark size=0.4*1.5*\mymarker,every mark/.append style={fill=MyBlue},mark=diamond*\\%
MyRed,only marks,mark size=1.2*\mymarker,every mark/.append style={fill=MyRed},mark=*\\%
MyRed,only marks,mark size=1.2*0.4*\mymarker,every mark/.append style={fill=MyRed},mark=*\\%
MyRed,only marks,mark size=1.2*0.4*\mymarker,every mark/.append style={fill=MyRed},mark=*\\%
}

\pgfplotscreateplotcyclelist{myadevcycle3}{%
black,only marks,mark size=\mymarker,every mark/.append style={fill=black},mark=square*\\%
MyRed,only marks,mark size=1.2*\mymarker,every mark/.append style={fill=MyRed},mark=*\\%
MyBlue,only marks,mark size=1.5*\mymarker,every mark/.append style={fill=MyBlue},mark=diamond*\\%
}

\pgfplotscreateplotcyclelist{myadevcyclewrap}{%
black,only marks,mark size=0.5*\mymarker,every mark/.append style={fill=black},mark=square*\\%
black,only marks,mark size=0.2*\mymarker,every mark/.append style={fill=black},mark=square*\\%
black,only marks,mark size=0.2*\mymarker,every mark/.append style={fill=black},mark=square*\\%
MyRed,only marks,mark size=0.6*\mymarker,every mark/.append style={fill=MyRed},mark=*\\%
MyRed,only marks,mark size=0.6*0.4*\mymarker,every mark/.append style={fill=MyRed},mark=*\\%
MyRed,only marks,mark size=0.6*0.4*\mymarker,every mark/.append style={fill=MyRed},mark=*\\%
MyBlue,only marks,mark size=0.75*\mymarker,every mark/.append style={fill=MyBlue},mark=diamond*\\%
MyBlue,only marks,mark size=0.75*0.4*\mymarker,every mark/.append style={fill=MyBlue},mark=diamond*\\%
MyBlue,only marks,mark size=0.75*0.4*\mymarker,every mark/.append style={fill=MyBlue},mark=diamond*\\%
}

\pgfplotscreateplotcyclelist{SME}{%
black,only marks,mark size=\mymarker,every mark/.append style={fill=black},mark=square*\\%
MyRed,only marks,mark size=2.3*\mymarker,every mark/.append style={fill=MyRed},mark=*\\%
}

\pgfkeys{
    /pgf/number format/.cd,
        set thousands separator={\,},
        min exponent for 1000 sep=4,
}

\newcommand{\mymarker}{2pt}
\newcommand{\myline}{thick}

\begin{document}

%\vspace*{4cm}
\title{Ground Tests of Einstein's Equivalence Principle: From Lab-based to 10-m Atomic Fountains}

\author{D. Schlippert}
\author{H. Albers}
\author{L. L. Richardson}
\author{D. Nath}
\author{H. Heine}
\author{C. Meiners}
\author{\mbox{{\'E}. Wodey}}
\author{A. Billon}
\author{J. Hartwig}
\author{C. Schubert}
\author{N. Gaaloul}
\author{W. Ertmer}
\author{E. M. Rasel}\email{rasel@iqo.uni-hannover.de}
\affiliation{Institut f\"ur Quantenoptik and Centre for Quantum Engineering and Space-Time Research \textnormal{(QUEST)}, Leibniz Universit\"at Hannover, Welfengarten 1, D-30167 Hannover, Germany}

\date{\today}% It is always \today, today,
             % but any date may be explicitly specified

\begin{abstract}
To date, no framework combining quantum field theory and general relativity and hence unifying all four fundamental interactions, exists. Violations of the \name{Einstein's} equivalence principle (EEP), being the foundation of general relativity, may hold the key to a theory of \quot{quantum gravity}. The universality of free fall (UFF), which is one of the three pillars of the EEP, has been extensively tested with classical bodies. Quantum tests of the UFF, e.g. by exploiting matter wave interferometry, allow for complementary sets of test masses, orders of magnitude larger test mass coherence lengths and investigation of spin-gravity coupling. 
We review our recent work towards highly sensitive matter wave tests of the UFF on ground. In this scope, the first quantum test of the UFF utilizing two different chemical elements, \Rb~and \K, yielding an \Eotvos~ratio \Eot~$=\etav$ has been performed. We assess systematic effects currently limiting the measurement at a level of parts in $10^8$ and finally present our strategies to improve the current state-of-the-art with a test comparing the free fall of rubidium and ytterbium in a very long baseline atom interferometry setup. Here, a $\SI{10}{m}$ baseline combined with a precise control of systematic effects will enable a determination of the \Eotvos~ratio at a level of parts in $10^{13}$ and beyond, thus reaching and overcoming the performance limit of the best classical tests.
\end{abstract}

%\pacs{03.75.Be, 03.75.Dg, 04.80.Cc, 06.30.Gv, 37.25.+k}% PACS, the Physics and Astronomy
                             % Classification Scheme.
%\keywords{Suggested keywords}%Use showkeys class option if keyword
                              %display desired
\maketitle

\section{Introduction}
%
%\subsection{Testing Einstein's equivalence principle}
%
With the great success of the grand unification theory~\cite{Georgi74PRL} the question arose whether the remaining fourth interaction, gravitation, could be unified with the other three yielding a \quot{theory of everything}. However, all approaches trying to merge quantum field theory and general relativity to a \quot{quantum gravity} framework consistent over all energy scales have failed so far~\cite{Laemmerzahl06APB}. Hence, in spite of both theories being confirmed at outstanding precision on their own, extensions of at least one of them, e.g. additional fields, are necessary in order to resolve their incompatibility.\\
General relativity is fully based on the postulates constituting \name{Einstein's} equivalence principle (EEP). Next to local position invariance and local \name{Lorentz} invariance, the EEP comprises the universality of free fall (UFF), which states that in absence of other forces all bodies located at the same space-time point experience the same acceleration in a gravitational field independently of their composition when neglecting self-gravity. While scrutinizing the EEP, it moreover was identified that under certain circumstances the UFF can be treated as direct empirical foundation for EEP~\cite{Will14}. Hence, tests of the UFF are a promising candidate in order to further investigate possible extensions of our understanding of gravity compatible with a theory of \quot{quantum gravity}.\\
A validity of the UFF implies the equality of inertial mass $m_{\text{in}}$ and gravitational mass $m_{\text{gr}}$ of any test body. In 1884, \name{Hertz} described the fact that gravity, unlike any other interaction, acts identically on all bodies independently of their gravitational charge as a \textit{\quot{wonderful mystery}}~\cite{Hertz99}. The so called \Eotvos~ratio
\begin{equation}\label{eq:eotvos}
\eta_{\,\text{A,B}}\equiv 2\thickspace\frac{g_{\text{A}}-g_{\text{B}}}{g_{\text{A}}+g_{\text{B}}}
=2\thickspace\frac{\left(\frac{m_{\text{gr}}}{m_{\text{in}}}\right)_{\text{A}}
-\left(\frac{m_{\text{gr}}}{m_{\text{in}}}\right)_{\text{B}}}
{\left(\frac{m_{\text{gr}}}{m_{\text{in}}}\right)_{\text{A}}
+\left(\frac{m_{\text{gr}}}{m_{\text{in}}}\right)_{\text{B}}}\;,
\end{equation}
where $g_{\text{i}}$ is the gravitational acceleration of test body $i=A,B$ is a comprehensive figure of merit when testing the UFF of test bodies A and B and is non-zero in case of a violation of the UFF.\\
Tests of the UFF emerged from \name{Galilei's} thought experiment in the 16\textsuperscript{th} century of comparing the free fall of different cannon balls dropped from the leaning tower of Pisa, commonly referred to as \name{Galilean} tests~\cite{Stillman03}. A demonstration test of this kind was performed during the Apollo 15 mission in 1971 by dropping a hammer and a feather on the Moon~\cite{Apollo15}. The most accurate measurements of the Eötvös ratio were performed by i) monitoring the distance between Earth and Moon in free fall around the Sun by means of laser ranging~\cite{Williams04PRL,Mueller12CQG}, yielding $\eta_{\,\text{Earth,Moon}}=(-0.8\pm 1.3)\times 10^{-13}$ and ii) employing a torsion balance~\cite{Eotvos89} with beryllium and titanium test masses~\cite{Schlamminger08PRL} yielding $\eta_{\,\text{Be,Ti}}=(0.3\pm 1.8)\times 10^{-13}$. The best \name{Galilean} test used a laser interferometer to read out the differential free fall motion of copper and uranium test masses~\cite{Niebauer87PRL} and found $\eta_{\,\text{Cu,U}}=(1.3\pm 5.0)\times 10^{-10}$.\\
%
%\subsection{Quantum tests of the universality of free fall}
%
The aforementioned tests employ classical, macroscopic bodies as test masses. In a complementary approach, the UFF can also be tested with quantum objects by observing the interference of massive particles such as neutrons or atoms under the influence of gravity. As first demonstrated in 1973 by \name{Colella}, \name{Overhauser}, and \name{Werner}~\cite{Colella75PRL}, the gravitationally induced phase shift imprinted on a particle's wave function is either compared to a classical gravimeter or to a second quantum object.\\
Quantum tests of the UFF differ from their classical counterparts in various aspects. Matter wave tests extend the set of test masses by allowing to employ any laser-coolable species. Furthermore, use of cold atoms add the spin as a degree of freedom and enables investigation of spin-gravity coupling~\cite{Laemmerzahl06APB}, and the accessible ultracold temperatures are inherently linked to macroscopic coherence lengths~\cite{Goeklue08CQG} which stands in fundamental contrast to classical test masses.\\ %purposely left out test duality --> STE-QUEST. add?
Quantum tests of the UFF that have been performed in the past can be classified in three categories: i) semi-classical tests, comparing an atom interferometer to a classical gravimeter~\cite{Peters99Nature,Merlet10Metrologia} and reaching accuracies on the ppb-level; ii) quantum tests at a level of parts in $10^7$ comparing the free fall of rubidium~\cite{Fray04PRL,Bonnin13PRA,Zhou15arxiv} or strontium~\cite{Tarallo14PRL} isotopes; iii) quantum tests comparing the free fall of two different chemical elements~\cite{Schlippert14PRL}.\\
%%
%\begin{table}[bt]
%\centering
%\small\renewcommand{\arraystretch}{1.4}
%\caption[]{Overview of performed matter wave tests of the UFF.}
%\label{tab:oldexperiments}
%\input{input/oldexperiments.tex}
%\end{table}
%%
Analyzing a test mass pair in a given framework, e.g. a test theory~\cite{Damour12CQG} or a parametrization~\cite{Hohensee13PRL}, allows to quantify the influence of a violation of the UFF ruled out with a given test mass pair. In general, a well-suited test mass pair fulfills $m_{\text{A}}\gg m_{\text{B}}$ or vice versa, making different chemical elements generally interesting test pairs. Accordingly, with their naturally low relative mass difference comparisons of heavier isotopes suffer from lower sensitivity to violations. On the other hand, however, they benefit from strong rejection of noise sources~\cite{Bonnin13PRA} and systematic errors~\cite{Aguilera14CQG,Hogan08arXiv}.\\\\ 
%SME plot?
This article is organized as follows: In \cref{sec:QTUFF}, we provide a brief overview on the underlying theory of dual species matter wave interferometry and summarize the first quantum test of the UFF using two different chemical elements, \Rb~and \K. We furthermore discuss an assessment of the systematic biases influencing our measurement. \Cref{sec:VLBAI} focuses on our strategies aiming towards a state-of-the-art test of the UFF comparing the free fall of ytterbium and rubidium in a $\SI{10}{m}$ very long base line atom interferometry setup. This article closes with an outlook into the future of matter wave tests of the UFF and a conclusion in \cref{sec:conclusion}.
\section{Quantum test of the universality of free fall of \Rb~and \K}\label{sec:QTUFF}
\begin{figure}[bt]
\begin{center}
\includegraphics[width=.4\linewidth]{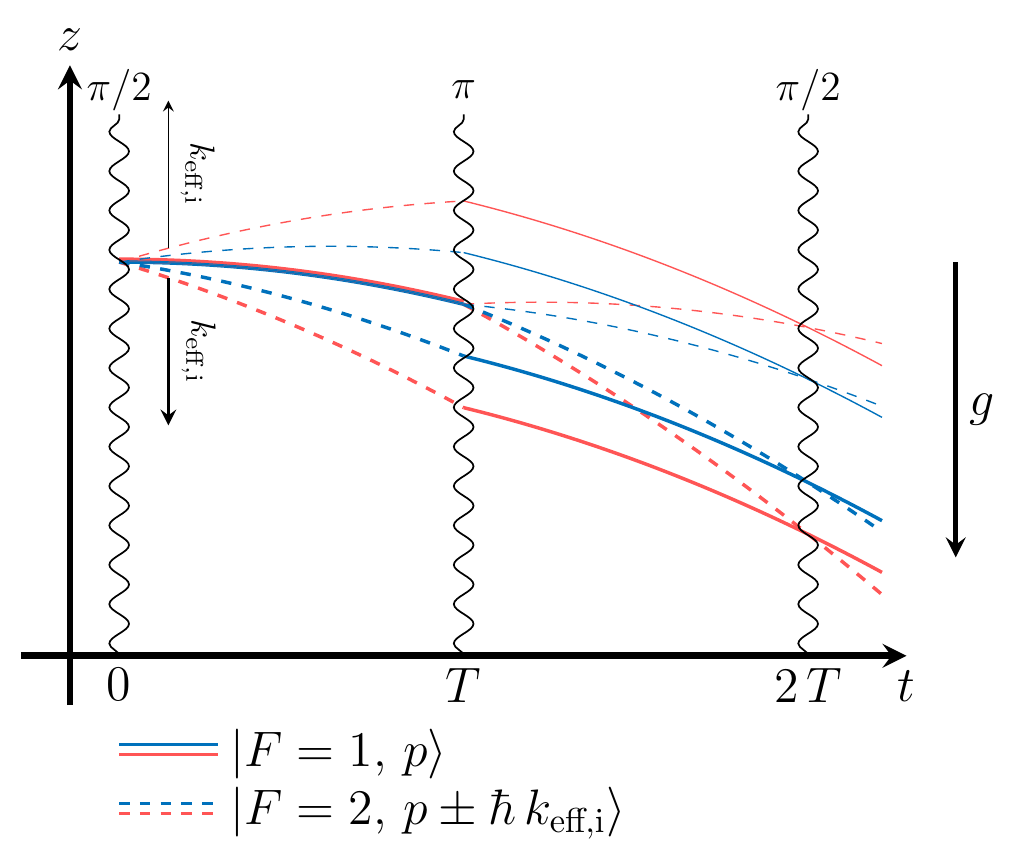}
\caption[]{Space-time diagram of a dual-species \name{Mach-Zehnder} matter wave interferometer in a constant gravitational field for the downward (thick lines) and upward (thin lines) direction of momentum transfer. Stimulated \name{Raman} transitions at times $0$, $T$, and $2\,T$ couple the states $\ket{F_i=1,\,p}$ and $\ket{F_i=2,\,p\pm\hbar\,\keffi{i}}$, where $i$ stands for Rb (blue lines) or K (red lines). The velocity change induced by the \name{Raman} pulses is not to scale with respect to the gravitational acceleration.}
\label{fig:mzdual}
\end{center}
\end{figure}

\begin{figure}[bt]
\includegraphics[width=.9\linewidth]{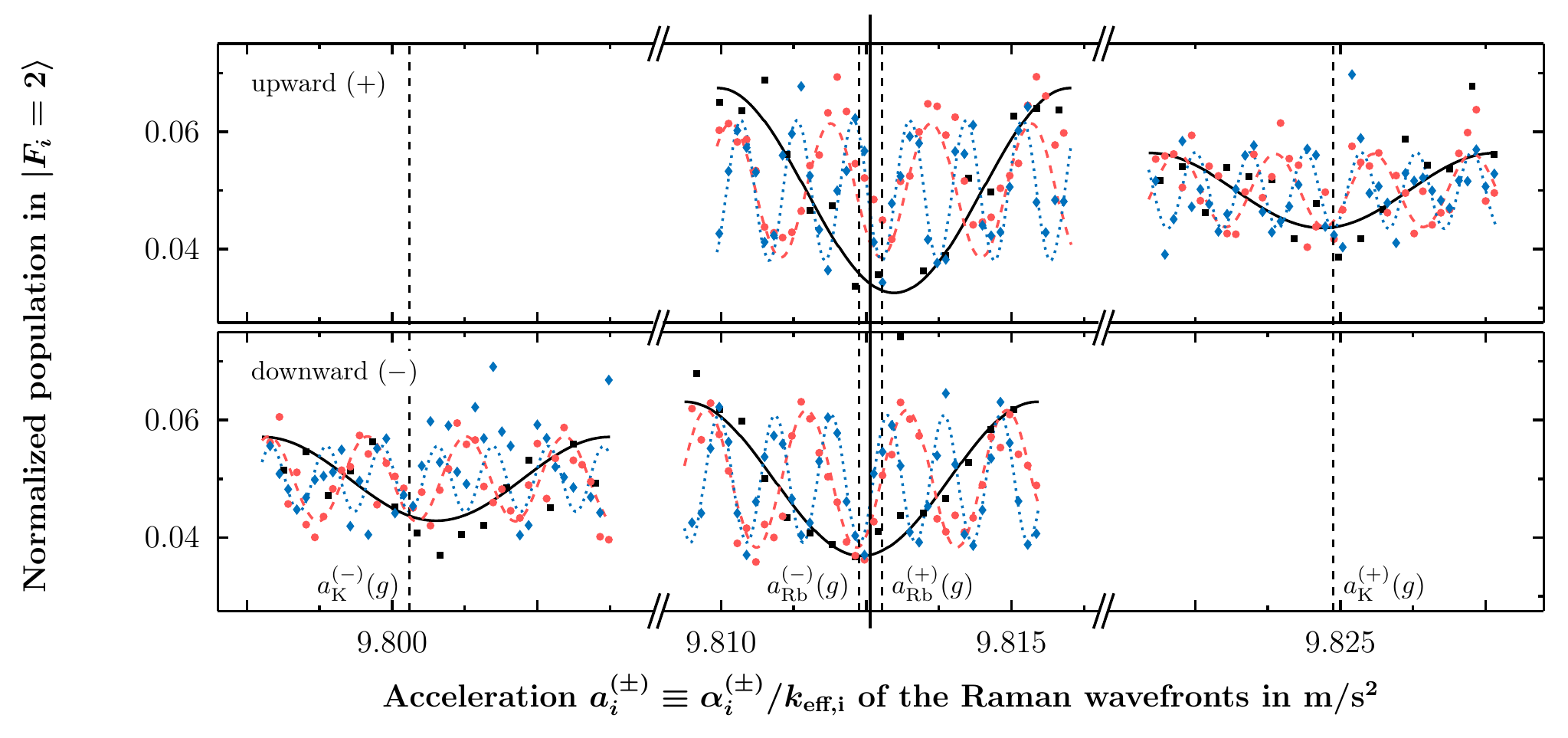}
\caption[Determination of the differential gravitational acceleration of rubidium and potassium]{Determination of the differential gravitational acceleration of rubidium and potassium. Typical fringe signals and sinusoidal fit functions are plotted in dependence of the effective \name{Raman} wavefront acceleration for pulse separation times $T=\SI{8}{ms}$ (black squares and solid black line), $T=\SI{15}{ms}$ (red circles and dashed red line), and $T=\SI{20}{ms}$ (blue diamonds and dotted blue line) for upward $(+)$ and downward $(-)$ direction of momentum transfer. The central fringe positions $\mathbf{a_i^{(\pm)}(g)}$ (dashed vertical lines), where $i$ is Rb or K, are shifted symmetrically around $g_i=[a_i^{(+)}(g)-a_i^{(-)}(g)]/2$ (solid vertical line). The data sets are corrected for slow linear drifts and offsets.
}
\label{fig:qtufffringes}
\end{figure}

In order to observe the gravitational acceleration acting on \Rb~and \K, we employ the \name{Mach-Zehnder}-type matter wave interferometer geometry~\cite{Kasevich91PRL} realized with stimulated \name{Raman} transitions coupling the states $\ket{F_i=1,\,p}$ and $\ket{F_i=2,\,p\pm\hbar\,\keffi{i}}$ as displayed in \cref{fig:mzdual}. In this configuration, we make use of an effective wavefront acceleration $\frac{\alpha}{\keff}$ caused by a linear frequency ramp $\alpha$ of the beam splitting light frequency difference with effective wave vector $\keff$. This acceleration enters the leading order phase shift as (throughout this Section, $i$ is Rb or K)
\begin{equation}\label{eq:phaseshift}
\Delta\phi_i=(g_i-\frac{\alpha_i}{\keffi{i}})\cdot\keffi{i}\cdot T^2\;.
\end{equation}
An experimental cycle starts by collecting $\SI{8e8}{atoms}$ of \Rb~and $\SI{3e7}{atoms}$ of \K~from a transversely cooled atomic beam within $\SI{1}{s}$ in a three-dimensional magneto-optical trap. The ensembles are subsequently cooled down to sub-\name{Doppler} temperatures utilizing the techniques described in Refs.~\cite{Landini11PRA,Chu98,Phillips98} yielding temperatures $T_{\text{Rb}}=\SI{27}{\micro K}$ and $T_{\text{K}}=\SI{32}{\micro K}$. Optical pumping is utilized to prepare the atoms in the $\ket{F_i=1}$ Zeeman manifold. By switching off all cooling light fields, the atoms are subsequently released into free fall.\\
A sequence of three Raman light pulses separated by the time $T$ is employed to form a \name{Mach-Zehnder}-type interferometer while applying a linear chirp $\alpha$ on the \name{Raman} laser difference frequency causing an acceleration of the wavefronts of the beam splitters. Afterwards, the exit ports of the interferometer are selectively read out by optical pumping and detection of fluorescence driving the $\ket{F_i=2}\rightarrow\ket{F'_i=3}$ transition. A single experimental cycle takes $\approx\SI{1.6}{s}$.\\
By varying the the effective wavefront acceleration, a global phase minimum appears independently of the free evolution time $T$ where $g-\frac{\alpha}{\keff}=0$ and thus allows to determine $g$. \Cref{fig:qtufffringes} shows the determination of gravitational acceleration $a_i^{(\pm)}(g)$ of \Rb~and \K~for the upward and downward direction of momentum transfer. Here, observation of the phase shift for both directions allows to strongly suppress systematic phase shifts that do not invert their sign when changing directions of momentum transfer by computing the half difference signal~\cite{McGuirk02PRA,Louchet-Chauvet11NJP}.
\subsection{Data analysis}
For testing the universality of free fall, the global phase minimum positions $a_i^{(\pm)}(g)$ in \cref{fig:qtufffringes} are monitored continuously over $\approx\SI{4}{h}$ by tuning the effective acceleration of the \name{Raman} wavefronts $\alpha_i^{(\pm)}/\keffi{i}$ around $a_i^{(\pm)}(g)$ in 10 steps per direction of momentum transfer with pulse separation time $T=\SI{20}{ms}$. Accordingly, the acquisition of $g_i=[a_i^{(+)}(g)-a_i^{(-)}(g)]/2$ takes $\SI{32}{s}$ in total and yields one data point for the \Eotvos~ratio (\cref{eq:eotvos}). The statistical uncertainty of the \Eotvos~ratio measurement after $\SI{4096}{s}$ of integration is $\sigma_\eta=\stat$, dominated by technical noise of the potassium interferometer.\\
In Table~\ref{tab:systematics} we list systematic effects influencing our measurement with overall bias of $\Delta\eta_{\text{tot}}=-5.4\times 10^{-8}$ and an uncertainty $\delta\eta_{\text{tot}}=3.1\times 10^{-8}$. A third column $\delta\eta^{\text{adv}}$ shows expected improved uncertainties at an overall level of parts per billion when using a dual-species optical dipole trap~\cite{Zaiser11PRA} as a common source which allows to precisely collocate the ensembles and to control their differential center of mass motion and expansion.
\begin{table}[bt]
\centering
\small\renewcommand{\arraystretch}{1.4}
\caption{Overview systematic biases  \EotDelta~and comparison of the current  uncertainties $\delta\eta$ of the \Eotvos~ratio to the improved uncertainties $\delta\eta^{\text{adv}}$ achieved by using an optical dipole trap. The uncertainties are treated to be uncorrelated at the level of inaccuracy.}
\label{tab:systematics}
\begin{center}
\begin{tabular}{|l|c|c|c|}
\hline
Contribution & \EotDelta &\Eotsigma & $\delta\eta^{\text{adv}}$ \\
\hline

Second-order \name{Zeeman} effect& $-5.8\times 10^{-8}$ &$2.6\times 10^{-8}$&$\mathbf{3.0\times 10^{-9}}$\\

Wavefront aberration & 0 &$1.2\times 10^{-8}$&$\mathbf{3.0\times 10^{-9}}$\\

\name{Coriolis} force& 0 &$9.1\times 10^{-9}$&$\mathbf{1.0\times 10^{-11}}$\\

Two-photon light shift& $4.1\times 10^{-9}$ &$8.2\times 10^{-11}$&$8.2\times 10^{-11}$\\

Effective wave vector& 0 &$1.3\times 10^{-9}$&$1.3\times 10^{-9}$\\

First-order gravity gradient& 0 &$9.5\times 10^{-11}$&$\mathbf{1.0\times 10^{-12}}$\\
\hline
Total &$-5.4\times 10^{-8}$&$3.1\times 10^{-8}$&$\mathbf{4.4\times 10^{-9}}$\\
\hline

\end{tabular}
\end{center}

\end{table}
\subsection{Summary}
Taking into account the statistical uncertainty $\sigma_\eta$ and the bias $\Delta\eta_{\text{tot}}$, the \Eotvos~ratio can be determined to \Eot~$=\etav$. At the current stage, the experiment is solely limited by technical noise dominating the short-term instability of the potassium interferometer. Hence, in the quadratic sum the statistical uncertainty fully overrules our systematic uncertainty. By reducing technical noise sources common mode noise rejection~\cite{Barrett15arxiv} between the interferometers will allow to push the experiment towards its limit posed by systematic uncertainty.
\section{Very long baseline atom interferometry}\label{sec:VLBAI}
\subsection{Experimental setup}
As shown in section~\ref{sec:QTUFF} for a \name{Mach-Zehnder}-type geometry, the sensitivity to accelerations of an atom interferometer scales with the square of the pulse separation time $T^2$.
A natural way to improve this sensitivity is to increase the free-fall time of the atoms enabling longer pulse separation times. This is the main driver for ground-based very long baseline devices and micro-gravity experiments. The latter feature free-fall times up to several seconds (droptower, parabolic flights), minutes (sounding rockets), or even days (space stations, satellites) in a small and thus well characterized volume. The practical and technological challenges combined with the high costs limit, however, the use of such platforms. In this section, we report about an on-going project of a ground-based very long baseline atom interferometer (VLBAI) device that will extend the baseline of the apparatus described in section~\ref{sec:QTUFF} from \SI{30}{\centi\meter} to more than \SI{10}{\meter}, allowing atoms to experience free-fall times up to $2T\sim\SI{1.3}{\second}$ in drop mode or up to $2T\sim\SI{2.6}{\second}$ in fountain mode. Together with our choice of species described below, we expect to reach an inaccuracy of $7\cdot 10^{-13}$ in the \Eotvos~ratio in the near future~\cite{Hartwig15NJP}.
\\As a device targeting a quantum test of the UFF, the proposed apparatus is designed as a dual-species gravimeter using ultra-cold mixtures of rubidium and ytterbium. The relevance of this species choice is motivated by the constraints possible to put on UFF violating theories, such as the dilaton scenario~\cite{Damour12CQG} and the standard model extension~\cite{Hohensee13PRL} (SME). In particular, an analysis in the SME framework shows that the Rb-Yb test pair choice is complementary to the Rb-K pair which was chosen for the previously described project, the QUANTUS/MAIUS/PRIMUS micro-gravity experiments~\cite{Rudolph15NJP,Seidel13proceedings,Herrmann12} and the STE-QUEST~\cite{Schubert13arxiv,Aguilera14CQG} M4 satellite proposal.
\\The extended size of the apparatus triggers specific engineering challenges to reach the UFF test performance announced above. As already demonstrated in other precision atom interferometers, a rotation compensation~\cite{Lan12PRL,Sugarbaker13PRL,Dickerson13PRL} of the inertial reference mirror at rates of $\sim\si{\micro\radian\per\second}$ is required in order to mitigate the systematic uncertainty linked to the \name{Coriolis} effect. Moreover, the use of rubidium atoms with magnetic susceptibility~\cite{Steck} of \SI{57.5}{\giga\hertz\per\tesla\squared} requires magnetic shielding of a factor at least \num{1e4} along the entire interferometry region. In this case, it extends over more than \SI{10}{\meter}. Finally, the reduced diameter of the vacuum tube (for efficient magnetic shielding) limits its conductance and makes its evacuation down to \SI{1e-10}{\milli\bar} challenging.\\

\subsection{Atomic sources}

In order to fully take advantage of the long baseline without severe systematics limiting the performance, the size of the clouds during their free-fall must be kept as small as possible. This can be achieved by delta-kick collimation (DKC) techniques~\cite{Chu1986pro,AmmanPRL1997} already demonstrated in the scope of micro-gravity experiments~\cite{Muentinga13PRL} or very-long-baseline atom fountains~\cite{KovachyPRL2015}. In the current design, we plan for a mixture of rubidium and ytterbium with $2\cdot 10^5$ and $1\cdot 10^5$ atoms, respectively. Preliminary estimations show that with a DKC pulsed at few tens of milliseconds, it is possible to keep the radius of the mixture at around $\SI{2}{mm}$ after $\SI{1.5}{s}$ of free evolution time. Within this regime, the leading systematics effects are not expected to deteriorate the uncertainty of the UFF test~\cite{Hartwig15NJP}. 
\\Furthermore, the preparation time of such an ultra-cold mixture should not exceed \SI{10}{\second} in order to enable sufficient repetition rates for reaching a statistical error of $7\cdot10^{-13}$ after one day of averaging. This cycling rates should be within reach in view of recent development in the production of high-flux sources of degenerate gases~\cite{Rudolph15NJP}.  

\subsection{Dual-species launch for precision tests}

The initial collocation and differential velocity of the two atomic clouds need to be kept small and very well characterized. Indeed, gravity gradients couple to the initial spatial offset and differential velocity inducing detrimental phase shifts at the output ports of the dual-interferometer~\cite{Hogan08arXiv}. More precisely, the desired accuracy for a UFF test implies a maximum offset between the two clouds of about \SI{10}{\nano\meter} and a maximum relative velocity of about \SI{10}{\nano\meter\per\second} derived in previous work~\cite{Hartwig15NJP}. Beyond these limits, the characterization of the gravity gradients becomes extremely challenging. 
In the condensed regime, the interactions play a crucial role in defining the symmetry of the ground state of the mixture. For a large overlap between the two test species, the choice of isotopes has to be restricted to miscible pairs. In a previous study~\cite{Hartwig15NJP}, we showed that the isotopes \Yb{168}~and \Yb{170}~can be good candidates to mix with \Rb thanks to their scattering length properties. Their natural abundances of $3\%$ and $0.1\%$, respectively, increase, however, the challenge for a high-flux source of suitable cold ytterbium atoms.
The collocation requirement implies the use of a common trap for both species. Since the ground state of bosonic ytterbium cannot be magnetically trapped, a mid-infrared dipole trap will be used for this purpose. 
In order to fully unfold the potential of the baseline in terms of achievable free fall time, a fountain launch is necessary. Due to the very small diffrerential velocity allowed here, molasses launch is not sufficiently accurate. In a recent proposal~\cite{Chamakhi2015}, it was shown that a single lattice cannot drive atoms with different masses to the same velocity after an acceleration ramp. The use of two lattices to control each species is not possible due to crosstalks between the atoms transitions and the two light frequencies. It was rather suggested in the latter proposal to utilize two lattices at \textit{tune-out} or \textit{zero-magic} frequencies of one atom each. For rubidium, light frequencies for which the contribution of the D$_1$ and D$_2$ lines to the dipole potential balance, were recently precisely measured~\cite{HeroldPRL2012} to an uncertainty below $\SI{1}{pm}$. Concerning ytterbium, there are, to our knowledge, no experimental data available but only theoretical calculations~\cite{ChengMitroyPRA2013} predicting tune-out wavelengths at $\SI{358.78}{nm}$ and $\SI{553.06}{nm}$ with a large uncertainty of a fraction of a nanometer. It is therefore highly interesting to experimentally determine these wavelength for fundamental as well as practical reasons. Once this is done, it is possible to engineer a selective lattice launch accelerating the two atomic species to equal velocities up to few nm/s as suggested~\cite{Chamakhi2015} for rubidium and potassium.
\\The baseline presented in this section, would in this case close the precision gap between classical and quantum UFF tests utilizing interferometers with free fall times of up to $\SI{2.8}{\second}$. 
\section{Outlook \& Conclusion}\label{sec:conclusion}
Matter wave interferometers are a new tool with fascinating prospects for future investigations of gravity, its relation to quantum mechanics and related open questions~\cite{Hamilton15PRL,Hamilton15arxiv}. We demonstrated a test of the UFF with the two different chemical elements Rb and K to a level of $5\cdot10^{-7}$. With the same apparatus we anticipate an improvement by two orders of magnitude with the implementation of an optical dipole trap. We are setting up a large scale experiment with increased free fall time, targetting a UFF test with Rb and Yb to the level of $7\cdot10^{-13}$ competitive with classical tests. Pursuing tests of the universality of free fall is a very promising strategy to find the missing piece for a self-consistent \quot{quantum gravity} framework valid over all energy scales and complementary to \name{Galilean} tests in space~\cite{Touboul12CQG}. Matter wave interferometry is not only enlarging the choice of test materials, but also allows to probe gravity with new states of matter such as entangled atoms or even Schrödinger cats.
\section*{Acknowledgments}
We are grateful to W. P. Schleich, A. Roura, and H. Ahlers for fruitful discussions.\\
This work was funded by the German Research Foundation (DFG) via the Sonderforschungsbereich (SFB) 1128 Relativistic Geodesy and Gravimetry with Quantum Sensors (geo-Q) and the Cluster of Excellence Centre for Quantum Engineering and Space-Time Research (QUEST). It is also supported by the German Space Agency (DLR) with funds provided by the Federal Ministry for Economic Affairs and Energy (BMWi) due to an enactment of the German Bundestag under grant numbers DLR 50WM1131-1137 (QUANTUS-III) and DLR 50WM1142 (PRIMUS-II).
\section*{References}
\bibliography{RASEL}

\begin{thebibliography}{10}

\bibitem{Georgi74PRL}
H.~Georgi and S.~L. Glashow.
\newblock Unity of all elementary-particle forces.
\newblock {\em Phys. Rev. Lett.}, 32(8):438--441, 1974.

\bibitem{Laemmerzahl06APB}
C.~L\"ammerzahl.
\newblock The search for quantum gravity effects i.
\newblock {\em Appl. Phys. B}, 84(4):551--562--, 2006.

\bibitem{Will14}
C.~M. Will.
\newblock The confrontation between general relativity and experiment.
\newblock {\em Living Rev. Relativity}, 17(4), 2014.

\bibitem{Hertz99}
H.~Hertz.
\newblock {\em Die Constitution der Materie: Eine Vorlesung \"uber die
  Grundlagen der Physik aus dem Jahre 1884}.
\newblock Springer, Berlin, 1999.

\bibitem{Stillman03}
D.~Stillman.
\newblock {\em Galileo at work : his scientific biography}.
\newblock Mineola (N.Y.): Dover publ., 2003.

\bibitem{Apollo15}
Manned Spacecraft~Center NASA.
\newblock Apollo 15 preliminary science report.
\newblock {\em NASA SP}, 289, 1972.

\bibitem{Williams04PRL}
J.~G. Williams, S.~G. Turyshev, and D.~H. Boggs.
\newblock Progress in lunar laser ranging tests of relativistic gravity.
\newblock {\em Phys. Rev. Lett.}, 93(26):261101--, 2004.

\bibitem{Mueller12CQG}
J.~M{\"u}ller, F.~Hofmann, and L.~Biskupek.
\newblock Testing various facets of the equivalence principle using lunar laser
  ranging.
\newblock {\em Classical and Quantum Gravity}, 29(18):184006--, 2012.

\bibitem{Eotvos89}
L.~E\"otv\"os.
\newblock \"uber die anziehung der erde auf verschiedene substanzen.
\newblock {\em Mathematische and naturwissenschaftliche Berichte aus Ungarn},
  8:65--, 1889.

\bibitem{Schlamminger08PRL}
S.~Schlamminger, K.-Y. Choi, T.~A. Wagner, J.~H. Gundlach, and E.~G.
  Adelberger.
\newblock Test of the equivalence principle using a rotating torsion balance.
\newblock {\em Phys. Rev. Lett.}, 100(4):041101--, 2008.

\bibitem{Niebauer87PRL}
T.~M. Niebauer, M.~P. McHugh, and J.~E. Faller.
\newblock Galilean test for the fifth force.
\newblock {\em Phys. Rev. Lett.}, 59:609--612, 1987.

\bibitem{Colella75PRL}
R.~Colella, A.~W. Overhauser, and S.~A. Werner.
\newblock Observation of gravitationally induced quantum interference.
\newblock {\em Phys. Rev. Lett.}, 34(23):1472--1474, 1975.

\bibitem{Goeklue08CQG}
E.~G\"okl\"u and C.~L\"ammerzahl.
\newblock Metric fluctuations and the weak equivalence principle.
\newblock {\em Classical Quantum Gravity}, 25(10):105012--, 2008.

\bibitem{Peters99Nature}
A.~Peters, K.-Y. Chung, and S.~Chu.
\newblock Measurement of gravitational acceleration by dropping atoms.
\newblock {\em Nature (London)}, 400(6747):849--852, 1999.

\bibitem{Merlet10Metrologia}
S.~Merlet, Q.~Bodart, N.~Malossi, A.~Landragin, F.~Pereira~Dos Santos,
  O.~Gitlein, and L.~Timmen.
\newblock Comparison between two mobile absolute gravimeters: optical versus
  atomic interferometers.
\newblock {\em Metrologia}, 47(4):L9--, 2010.

\bibitem{Fray04PRL}
S.~Fray, C.~A. Diez, T.~W. H\"ansch, and M.~Weitz.
\newblock Atomic interferometer with amplitude gratings of light and its
  applications to atom based tests of the equivalence principle.
\newblock {\em Phys. Rev. Lett.}, 93(24):240404--, 2004.

\bibitem{Bonnin13PRA}
A.~Bonnin, N.~Zahzam, Y.~Bidel, and A.~Bresson.
\newblock Simultaneous dual-species matter-wave accelerometer.
\newblock {\em Phys. Rev. A}, 88(4):043615--, 2013.

\bibitem{Zhou15arxiv}
L.~Zhou, S.~Long, B.~Tang, X.~Chen, F.~Gao, W.~Peng, W.~Duan, J.~Zhong,
  Z.~Xiong, J.~Wang, Y.~Zhang, and M.~Zhan.
\newblock Test of equivalence principle at $10^{-8}$ level by a dual-species
  double-diffraction raman atom interferometer.
\newblock arXiv:1503.00401.

\bibitem{Tarallo14PRL}
M.~G. Tarallo, T.~Mazzoni, N.~Poli, D.~V. Sutyrin, X.~Zhang, and G.~M. Tino.
\newblock Test of einstein equivalence principle for 0-spin and
  half-integer-spin atoms: Search for spin-gravity coupling effects.
\newblock {\em Phys. Rev. Lett.}, 113(2):023005--, 2014.

\bibitem{Schlippert14PRL}
D.~Schlippert, J.~Hartwig, H.~Albers, L.~L. Richardson, C.~Schubert, A.~Roura,
  W.~P. Schleich, W.~Ertmer, and E.~M. Rasel.
\newblock Quantum test of the universality of free fall.
\newblock {\em Phys. Rev. Lett.}, 112:203002, 2014.

\bibitem{Damour12CQG}
T.~Damour.
\newblock Theoretical aspects of the equivalence principle.
\newblock {\em Classical Quantum Gravity}, 29(18):184001--, 2012.

\bibitem{Hohensee13PRL}
M.~A. Hohensee, H.~M\"uller, and R.~B. Wiringa.
\newblock Equivalence principle and bound kinetic energy.
\newblock {\em Phys. Rev. Lett.}, 111(15):151102--, 2013.

\bibitem{Aguilera14CQG}
D.~Aguilera et~al.
\newblock Ste-quest - test of the universality of free fall using cold atom
  interferometry.
\newblock {\em Classical and Quantum Gravity}, 31(11):115010--, 2014.

\bibitem{Hogan08arXiv}
J.~M. Hogan, D.~M.~S. Johnson, and M.~A. Kasevich.
\newblock Light-pulse atom interferometry.
\newblock arXiv:0806.3261.

\bibitem{Kasevich91PRL}
M.~Kasevich and S.~Chu.
\newblock Atomic interferometry using stimulated raman transitions.
\newblock {\em Phys. Rev. Lett.}, 67(2):181--184, 1991.

\bibitem{Landini11PRA}
M.~Landini, S.~Roy, L.~Carcagn\'i, D.~Trypogeorgos, M.~Fattori, M.~Inguscio,
  and G.~Modugno.
\newblock Sub-doppler laser cooling of potassium atoms.
\newblock {\em Phys. Rev. A}, 84(4):043432--, 2011.

\bibitem{Chu98}
S.~Chu.
\newblock Nobel lecture: The manipulation of neutral particles.
\newblock {\em Rev. Mod. Phys.}, 70(3):685--706, 1998.

\bibitem{Phillips98}
W.~D. Phillips.
\newblock {Nobel Lecture: Laser cooling and trapping of neutral atoms}.
\newblock {\em Rev. Mod. Phys.}, 70(3):721--741, Jul 1998.

\bibitem{McGuirk02PRA}
J.~M. McGuirk, G.~T. Foster, J.~B. Fixler, M.~J. Snadden, and M.~A. Kasevich.
\newblock Sensitive absolute-gravity gradiometry using atom interferometry.
\newblock {\em Phys. Rev. A}, 65(3):033608--, 2002.

\bibitem{Louchet-Chauvet11NJP}
A.~Louchet-Chauvet, T.~Farah, Q.~Bodart, A.~Clairon, A.~Landragin, S.~Merlet,
  and F.~Pereira Dos~Santos.
\newblock The influence of transverse motion within an atomic gravimeter.
\newblock {\em New J. Phys.}, 13(6):065025--, 2011.

\bibitem{Zaiser11PRA}
M.~Zaiser, J.~Hartwig, D.~Schlippert, U.~Velte, N.~Winter, V.~Lebedev,
  W.~Ertmer, and E.~M. Rasel.
\newblock Simple method for generating bose-einstein condensates in a weak
  hybrid trap.
\newblock {\em Phys. Rev. A}, 83:035601, 2011.

\bibitem{Barrett15arxiv}
B.~Barrett, L.~Antoni-Micollier, L.~Chichet, B.~Battelier, P.-A. Gominet,
  A.~Bertoldi, P.~Bouyer, and A.~Landragin.
\newblock Correlative methods for dual-species quantum tests of the weak
  equivalence principle.
\newblock arXiv:1503.08423.

\bibitem{Hartwig15NJP}
J.~Hartwig, S.~Abend, C.~Schubert, D.~Schlippert, H.~Ahlers, K.~Posso-Trujillo,
  N.~Gaaloul, W.~Ertmer, and E.~M. Rasel.
\newblock Testing the universality of free fall with rubidium and ytterbium in
  a very large baseline atom interferometer.
\newblock {\em New J. Phys.}, 17(3):035011--, 2015.

\bibitem{Rudolph15NJP}
J.~Rudolph et~al.
\newblock A high-flux bec source for mobile atom interferometers.
\newblock {\em New Journal of Physics}, 17(6):065001--, 2015.

\bibitem{Seidel13proceedings}
S.~T. Seidel, N.~Gaaloul, and E.~M. Rasel.
\newblock Maius - a rocket-born test of an atom interferometer with a
  chip-based atom laser.
\newblock {\em Proceedings of the 63rd International Astronautical Congress
  2012}, 3:801, 2013.

\bibitem{Herrmann12}
S.~Herrmann, H.~Dittus, and C.~L\"ammerzahl.
\newblock Testing the equivalence principle with atomic interferometry.
\newblock {\em Classical and Quantum Gravity}, 29(18):184003--, 2012.

\bibitem{Schubert13arxiv}
C.~Schubert et~al.
\newblock Differential atom interferometry with \textsuperscript{87}rb and
  \textsuperscript{85}rb for testing the uff in ste-quest.
\newblock arXiv:1312.5963.

\bibitem{Lan12PRL}
S.-Y. Lan, P.-C. Kuan, B.~Estey, P.~Haslinger, and H.~M\"uller.
\newblock Influence of the coriolis force in atom interferometry.
\newblock {\em Phys. Rev. Lett.}, 108(9):090402--, 2012.

\bibitem{Sugarbaker13PRL}
A.~Sugarbaker, S.~M. Dickerson, J.~M. Hogan, D.~M.~S. Johnson, and M.~A.
  Kasevich.
\newblock Enhanced atom interferometer readout through the application of phase
  shear.
\newblock {\em Phys. Rev. Lett.}, 111(11):113002--, 2013.

\bibitem{Dickerson13PRL}
S.~M. Dickerson, J.~M. Hogan, A.~Sugarbaker, D.~M.~S. Johnson, and M.~A.
  Kasevich.
\newblock Multiaxis inertial sensing with long-time point source atom
  interferometry.
\newblock {\em Phys. Rev. Lett.}, 111(8):083001--, 2013.

\bibitem{Steck}
D.~A. Steck.
\newblock {Rubidium 87 D Line Data, rev. 2.1.4}.
\newblock 2010.

\bibitem{Chu1986pro}
S~Chu, A~Ashkin Bjorkholm, P~Gordon, and LW~Hollberg.
\newblock Proposal for optically cooling atoms to temperatures of the order of
  10$^{-6}$\,{K}.
\newblock {\em Opt. Lett.}, 11:73, 1986.

\bibitem{AmmanPRL1997}
Hubert Ammann and Nelson Christensen.
\newblock Delta kick cooling: A new method for cooling atoms.
\newblock {\em Phys. Rev. Lett.}, 78:2088--2091, Mar 1997.

\bibitem{Muentinga13PRL}
H.~M\"untinga, H.~Ahlers, M.~Krutzik, A.~Wenzlawski, S.~Arnold, D.~Becker,
  K.~Bongs, H.~Dittus, H.~Duncker, N.~Gaaloul, C.~Gherasim, E.~Giese,
  C.~Grzeschik, T.~W. H\"ansch, O.~Hellmig, W.~Herr, S.~Herrmann, E.~Kajari,
  S.~Kleinert, C.~L\"ammerzahl, W.~Lewoczko-Adamczyk, J.~Malcolm, N.~Meyer,
  R.~Nolte, A.~Peters, M.~Popp, J.~Reichel, A.~Roura, J.~Rudolph,
  M.~Schiemangk, M.~Schneider, S.~T. Seidel, K.~Sengstock, V.~Tamma,
  T.~Valenzuela, A.~Vogel, R.~Walser, T.~Wendrich, P.~Windpassinger, W.~Zeller,
  T.~van Zoest, W.~Ertmer, W.~P. Schleich, and E.~M. Rasel.
\newblock Interferometry with bose-einstein condensates in microgravity.
\newblock {\em Phys. Rev. Lett.}, 110:093602, Feb 2013.

\bibitem{KovachyPRL2015}
Tim Kovachy, Jason~M. Hogan, Alex Sugarbaker, Susannah~M. Dickerson,
  Christine~A. Donnelly, Chris Overstreet, and Mark~A. Kasevich.
\newblock Matter wave lensing to picokelvin temperatures.
\newblock {\em Phys. Rev. Lett.}, 114:143004, Apr 2015.

\bibitem{Chamakhi2015}
R.~Chamakhi.
\newblock {Species-selective lattice launch for precision atom interferometry}.
\newblock {\em to be published}, 2015.

\bibitem{HeroldPRL2012}
C.~D. Herold, V.~D. Vaidya, X.~Li, S.~L. Rolston, J.~V. Porto, and M.~S.
  Safronova.
\newblock Precision measurement of transition matrix elements via light shift
  cancellation.
\newblock {\em Phys. Rev. Lett.}, 109:243003, Dec 2012.

\bibitem{ChengMitroyPRA2013}
Yongjun Cheng, Jun Jiang, and J.~Mitroy.
\newblock Tune-out wavelengths for the alkaline-earth-metal atoms.
\newblock {\em Phys. Rev. A}, 88:022511, Aug 2013.

\bibitem{Hamilton15PRL}
P.~Hamilton, M.~Jaffe, J.~M. Brown, L.~Maisenbacher, B.~Estey, and H.~M\"uller.
\newblock Atom interferometry in an optical cavity.
\newblock {\em Phys. Rev. Lett.}, 114(10):100405--, March 2015.

\bibitem{Hamilton15arxiv}
P.~Hamilton, M.~Jaffe, P.~Haslinger, Q.~Simmons, H.~M\"uller, and J.~Khoury.
\newblock Atom-interferometry constraints on dark energy.
\newblock arXiv:1502.03888.

\bibitem{Touboul12CQG}
P.~Touboul, G.~M\'etris, V.~Lebat, and A.~Robert.
\newblock The microscope experiment, ready for the in-orbit test of the
  equivalence principle.
\newblock {\em Classical Quantum Gravity}, 29(18):184010--, 2012.

\end{thebibliography}

\end{document}